\PassOptionsToPackage{unicode}{hyperref}
\PassOptionsToPackage{hyphens}{url}
\PassOptionsToPackage{dvipsnames,svgnames,x11names}{xcolor}
\documentclass[
]{article}
\usepackage{amsmath,amssymb}
\usepackage{lmodern}
\usepackage{iftex}
\ifPDFTeX
  \usepackage[T1]{fontenc}
  \usepackage[utf8]{inputenc}
  \usepackage{textcomp} % provide euro and other symbols
\else % if luatex or xetex
  \usepackage{unicode-math}
  \defaultfontfeatures{Scale=MatchLowercase}
  \defaultfontfeatures[\rmfamily]{Ligatures=TeX,Scale=1}
\fi
\IfFileExists{upquote.sty}{\usepackage{upquote}}{}
\IfFileExists{microtype.sty}{% use microtype if available
  \usepackage[]{microtype}
  \UseMicrotypeSet[protrusion]{basicmath} % disable protrusion for tt fonts
}{}
\makeatletter
\@ifundefined{KOMAClassName}{% if non-KOMA class
  \IfFileExists{parskip.sty}{%
    \usepackage{parskip}
  }{% else
    \setlength{\parindent}{0pt}
    \setlength{\parskip}{6pt plus 2pt minus 1pt}}
}{% if KOMA class
  \KOMAoptions{parskip=half}}
\makeatother
\usepackage{xcolor}
\usepackage{color}
\usepackage{fancyvrb}

\DefineVerbatimEnvironment{Highlighting}{Verbatim}{commandchars=\\\{\}}
\newenvironment{Shaded}{}{}

\newcommand{\DecValTok}[1]{\textcolor[rgb]{0.25,0.63,0.44}{#1}}

\newcommand{\ImportTok}[1]{\textcolor[rgb]{0.00,0.50,0.00}{\textbf{#1}}}

\newcommand{\KeywordTok}[1]{\textcolor[rgb]{0.00,0.44,0.13}{\textbf{#1}}}
\newcommand{\NormalTok}[1]{#1}
\newcommand{\OperatorTok}[1]{\textcolor[rgb]{0.40,0.40,0.40}{#1}}

\newcommand{\PreprocessorTok}[1]{\textcolor[rgb]{0.74,0.48,0.00}{#1}}

\newcommand{\StringTok}[1]{\textcolor[rgb]{0.25,0.44,0.63}{#1}}

\usepackage{longtable,booktabs,array}
\usepackage{calc} % for calculating minipage widths
\usepackage{etoolbox}
\makeatletter
\patchcmd\longtable{\par}{\if@noskipsec\mbox{}\fi\par}{}{}
\makeatother
\IfFileExists{footnotehyper.sty}{\usepackage{footnotehyper}}{\usepackage{footnote}}
\makesavenoteenv{longtable}
\usepackage{graphicx}
\makeatletter
\def\maxwidth{\ifdim\Gin@nat@width>\linewidth\linewidth\else\Gin@nat@width\fi}
\def\maxheight{\ifdim\Gin@nat@height>\textheight\textheight\else\Gin@nat@height\fi}
\makeatother
\setkeys{Gin}{width=\maxwidth,height=\maxheight,keepaspectratio}
\makeatletter
\def\fps@figure{htbp}
\makeatother
\providecommand{\tightlist}{%
  \setlength{\itemsep}{0pt}\setlength{\parskip}{0pt}}
\NewDocumentCommand\citeproctext{}{}
\NewDocumentCommand\citeproc{mm}{%
  \begingroup\def\citeproctext{#2}\cite{#1}\endgroup}
\makeatletter
 \let\@cite@ofmt\@firstofone
 \def\@biblabel#1{}
 \def\@cite#1#2{{#1\if@tempswa , #2\fi}}
\makeatother
\newlength{\cslhangindent}
\newlength{\csllabelwidth}
\newenvironment{CSLReferences}[2] % #1 hanging-indent, #2 entry-spacing
 {\begin{list}{}{%
  \setlength{\itemindent}{0pt}
  \setlength{\leftmargin}{0pt}
  \setlength{\parsep}{0pt}
  \ifodd #1
   \setlength{\leftmargin}{\cslhangindent}
   \setlength{\itemindent}{-1\cslhangindent}
  \fi
  \setlength{\itemsep}{#2\baselineskip}}}
 {\end{list}}
\usepackage{calc}

\ifLuaTeX
\usepackage[bidi=basic]{babel}
\else
\usepackage[bidi=default]{babel}
\fi
\babelprovide[main,import]{american}
\def\languageshorthands#1{}
\ifLuaTeX
  \usepackage{selnolig}  % disable illegal ligatures
\fi
\IfFileExists{bookmark.sty}{\usepackage{bookmark}}{\usepackage{hyperref}}
\IfFileExists{xurl.sty}{\usepackage{xurl}}{} % add URL line breaks if available
\hypersetup{
  pdftitle={Eunoia: Area-Proportional Euler and Venn Diagrams in Rust,
Julia, Python, R, and JavaScript},
  pdfauthor={Johan Larsson, Nicole M. Lee},
  pdflang={en-US},
  colorlinks=true,
  linkcolor={Maroon},
  filecolor={Maroon},
  citecolor={Blue},
  urlcolor={Blue},
  pdfcreator={LaTeX via pandoc}}

\title{Eunoia: Area-Proportional Euler and Venn Diagrams in Rust, Julia,
Python, R, and JavaScript}

\definecolor{c53baa1}{RGB}{83,186,161}
\definecolor{c202826}{RGB}{32,40,38}
\def \rorglobalscale {0.1}
\newcommand{\rorlogo}{%
\begin{tikzpicture}[y=1cm, x=1cm, yscale=\rorglobalscale,xscale=\rorglobalscale, every node/.append style={scale=\rorglobalscale}, inner sep=0pt, outer sep=0pt]
  \begin{scope}[even odd rule,line join=round,miter limit=2.0,shift={(-0.025, 0.0216)}]
    \path[fill=c53baa1,nonzero rule,line join=round,miter limit=2.0] (1.8164, 3.012) -- (1.4954, 2.5204) -- (1.1742, 3.012) -- (1.8164, 3.012) -- cycle;
    \path[fill=c53baa1,nonzero rule,line join=round,miter limit=2.0] (3.1594, 3.012) -- (2.8385, 2.5204) -- (2.5172, 3.012) -- (3.1594, 3.012) -- cycle;
    \path[fill=c53baa1,nonzero rule,line join=round,miter limit=2.0] (1.1742, 0.0669) -- (1.4954, 0.5588) -- (1.8164, 0.0669) -- (1.1742, 0.0669) -- cycle;
    \path[fill=c53baa1,nonzero rule,line join=round,miter limit=2.0] (2.5172, 0.0669) -- (2.8385, 0.5588) -- (3.1594, 0.0669) -- (2.5172, 0.0669) -- cycle;
    \path[fill=c202826,nonzero rule,line join=round,miter limit=2.0] (3.8505, 1.4364).. controls (3.9643, 1.4576) and (4.0508, 1.5081) .. (4.1098, 1.5878).. controls (4.169, 1.6674) and (4.1984, 1.7642) .. (4.1984, 1.8777).. controls (4.1984, 1.9719) and (4.182, 2.0503) .. (4.1495, 2.1132).. controls (4.1169, 2.1762) and (4.0727, 2.2262) .. (4.0174, 2.2635).. controls (3.9621, 2.3006) and (3.8976, 2.3273) .. (3.824, 2.3432).. controls (3.7505, 2.359) and (3.6727, 2.367) .. (3.5909, 2.367) -- (2.9676, 2.367) -- (2.9676, 1.8688).. controls (2.9625, 1.8833) and (2.9572, 1.8976) .. (2.9514, 1.9119).. controls (2.9083, 2.0164) and (2.848, 2.1056) .. (2.7705, 2.1791).. controls (2.6929, 2.2527) and (2.6014, 2.3093) .. (2.495, 2.3487).. controls (2.3889, 2.3881) and (2.2728, 2.408) .. (2.1468, 2.408).. controls (2.0209, 2.408) and (1.905, 2.3881) .. (1.7986, 2.3487).. controls (1.6925, 2.3093) and (1.6007, 2.2527) .. (1.5232, 2.1791).. controls (1.4539, 2.1132) and (1.3983, 2.0346) .. (1.3565, 1.9436).. controls (1.3504, 2.009) and (1.3351, 2.0656) .. (1.3105, 2.1132).. controls (1.2779, 2.1762) and (1.2338, 2.2262) .. (1.1785, 2.2635).. controls (1.1232, 2.3006) and (1.0586, 2.3273) .. (0.985, 2.3432).. controls (0.9115, 2.359) and (0.8337, 2.367) .. (0.7519, 2.367) -- (0.1289, 2.367) -- (0.1289, 0.7562) -- (0.4837, 0.7562) -- (0.4837, 1.4002) -- (0.6588, 1.4002) -- (0.9956, 0.7562) -- (1.4211, 0.7562) -- (1.0118, 1.4364).. controls (1.1255, 1.4576) and (1.2121, 1.5081) .. (1.2711, 1.5878).. controls (1.2737, 1.5915) and (1.2761, 1.5954) .. (1.2787, 1.5991).. controls (1.2782, 1.5867) and (1.2779, 1.5743) .. (1.2779, 1.5616).. controls (1.2779, 1.4327) and (1.2996, 1.3158) .. (1.3428, 1.2113).. controls (1.3859, 1.1068) and (1.4462, 1.0176) .. (1.5237, 0.944).. controls (1.601, 0.8705) and (1.6928, 0.8139) .. (1.7992, 0.7744).. controls (1.9053, 0.735) and (2.0214, 0.7152) .. (2.1474, 0.7152).. controls (2.2733, 0.7152) and (2.3892, 0.735) .. (2.4956, 0.7744).. controls (2.6016, 0.8139) and (2.6935, 0.8705) .. (2.771, 0.944).. controls (2.8482, 1.0176) and (2.9086, 1.1068) .. (2.952, 1.2113).. controls (2.9578, 1.2253) and (2.9631, 1.2398) .. (2.9681, 1.2544) -- (2.9681, 0.7562) -- (3.3229, 0.7562) -- (3.3229, 1.4002) -- (3.4981, 1.4002) -- (3.8349, 0.7562) -- (4.2603, 0.7562) -- (3.8505, 1.4364) -- cycle(0.9628, 1.7777).. controls (0.9438, 1.7534) and (0.92, 1.7357) .. (0.8911, 1.7243).. controls (0.8623, 1.7129) and (0.83, 1.706) .. (0.7945, 1.7039).. controls (0.7588, 1.7015) and (0.7252, 1.7005) .. (0.6932, 1.7005) -- (0.4839, 1.7005) -- (0.4839, 2.0667) -- (0.716, 2.0667).. controls (0.7477, 2.0667) and (0.7805, 2.0643) .. (0.8139, 2.0598).. controls (0.8472, 2.0553) and (0.8768, 2.0466) .. (0.9025, 2.0336).. controls (0.9282, 2.0206) and (0.9496, 2.0021) .. (0.9663, 1.9778).. controls (0.9829, 1.9534) and (0.9914, 1.9209) .. (0.9914, 1.8799).. controls (0.9914, 1.8362) and (0.9819, 1.8021) .. (0.9628, 1.7777) -- cycle(2.6125, 1.3533).. controls (2.5889, 1.2904) and (2.5553, 1.2359) .. (2.5112, 1.1896).. controls (2.4672, 1.1433) and (2.4146, 1.1073) .. (2.3529, 1.0814).. controls (2.2916, 1.0554) and (2.2228, 1.0427) .. (2.1471, 1.0427).. controls (2.0712, 1.0427) and (2.0026, 1.0557) .. (1.9412, 1.0814).. controls (1.8799, 1.107) and (1.8272, 1.1433) .. (1.783, 1.1896).. controls (1.7391, 1.2359) and (1.7052, 1.2904) .. (1.6817, 1.3533).. controls (1.6581, 1.4163) and (1.6465, 1.4856) .. (1.6465, 1.5616).. controls (1.6465, 1.6359) and (1.6581, 1.705) .. (1.6817, 1.7687).. controls (1.7052, 1.8325) and (1.7388, 1.8873) .. (1.783, 1.9336).. controls (1.8269, 1.9799) and (1.8796, 2.0159) .. (1.9412, 2.0418).. controls (2.0026, 2.0675) and (2.0712, 2.0804) .. (2.1471, 2.0804).. controls (2.223, 2.0804) and (2.2916, 2.0675) .. (2.3529, 2.0418).. controls (2.4143, 2.0161) and (2.467, 1.9799) .. (2.5112, 1.9336).. controls (2.5551, 1.8873) and (2.5889, 1.8322) .. (2.6125, 1.7687).. controls (2.636, 1.705) and (2.6477, 1.6359) .. (2.6477, 1.5616).. controls (2.6477, 1.4856) and (2.636, 1.4163) .. (2.6125, 1.3533) -- cycle(3.8015, 1.7777).. controls (3.7825, 1.7534) and (3.7587, 1.7357) .. (3.7298, 1.7243).. controls (3.701, 1.7129) and (3.6687, 1.706) .. (3.6333, 1.7039).. controls (3.5975, 1.7015) and (3.5639, 1.7005) .. (3.5319, 1.7005) -- (3.3226, 1.7005) -- (3.3226, 2.0667) -- (3.5547, 2.0667).. controls (3.5864, 2.0667) and (3.6192, 2.0643) .. (3.6526, 2.0598).. controls (3.6859, 2.0553) and (3.7155, 2.0466) .. (3.7412, 2.0336).. controls (3.7669, 2.0206) and (3.7883, 2.0021) .. (3.805, 1.9778).. controls (3.8216, 1.9534) and (3.8301, 1.9209) .. (3.8301, 1.8799).. controls (3.8301, 1.8362) and (3.8206, 1.8021) .. (3.8015, 1.7777) -- cycle;
  \end{scope}
\end{tikzpicture}
}

\usepackage[affil-it]{authblk}
\usepackage{orcidlink}
\author[1%
  ]{Johan Larsson%
    \,\orcidlink{0000-0002-4029-5945}\,%
    }
\author[2%
  ]{Nicole M. Lee%
    \,\orcidlink{0000-0001-6953-2177}\,%
    }

\affil[1]{Department of Mathematical Sciences, University of Copenhagen,
Denmark%
    \,\protect\href{https://ror.org/035b05819}{\protect\rorlogo}\,%
  }
\affil[2]{Program in Neuroscience and Behavioral Disorders, Duke-NUS
Medical School, Singapore%
    \,\protect\href{https://ror.org/02j1m6098}{\protect\rorlogo}\,%
  }
\date{30 October 2026}

\begin{document}
\maketitle

\section{Summary}\label{summary}

Eunoia is a \href{https://github.com/jolars/eunoia}{Rust
library}~(\citeproc{ref-eunoia2026}{Larsson, 2026}) for
area-proportional Euler and Venn diagrams. Given the sizes of a
collection of sets and their intersections, it fits a diagram of
circles, ellipses, squares, or rectangles whose overlapping areas match
those quantities as closely as possible. When an exact representation
does not exist, Eunoia returns the fitted layout with residuals and
goodness-of-fit statistics that help users assess how well it represents
the data. The Rust crate also powers an
\href{https://cran.r-project.org/package=eulerr}{R package (eulerr)}, a
\href{https://pypi.org/project/eunoia/}{Python package}, a
\href{https://github.com/jolars/Eunoia.jl}{Julia package}, a
\href{https://www.npmjs.com/package/@jolars/eunoia}{JavaScript package}
compiled to WebAssembly, a C API for further bindings, and a
\href{https://eunoia.bz/app}{web app} built on the JavaScript package.
All bindings call the same implementation and produce the same layout
given the same specification, options, and seed.

\section{Statement of Need}\label{statement-of-need}

Euler diagrams~(\citeproc{ref-euler1802}{Euler, 1802}) depict
relationships between sets, and area-proportional Euler diagrams do so
quantitatively: the area of each region is proportional to the size of
the corresponding set intersection. They are common in the life
sciences, for instance, where researchers use them to compare gene and
protein lists across conditions or studies. For most specifications
involving three or more sets, however, no exact area-proportional
diagram exists~(\citeproc{ref-wilkinson2012}{Wilkinson, 2012}), and the
diagram must instead be constructed
numerically~(\citeproc{ref-chow2007}{Chow, 2007}). Fitting a diagram is
therefore an optimization problem: choosing the shapes' positions and
parameters to minimize the discrepancy between fitted and requested
intersection areas.

Existing alternatives offer different compromises. venneuler introduced
multidimensional scaling~(MDS) initialization followed by overlap-aware
refinement~(\citeproc{ref-wilkinson2012}{Wilkinson, 2012}). venn.js
improved that initialization and added exact area
calculations~(\citeproc{ref-frederickson2015}{Frederickson, 2015}).
eulerAPE introduced ellipses for three-set
diagrams~(\citeproc{ref-micallef2014}{Micallef \& Rodgers, 2014}), and
eulerr generalized them to any number of
sets~(\citeproc{ref-larsson2018}{Larsson \& Gustafsson, 2018}).

Eunoia extends eulerr by

\begin{itemize}
\tightlist
\item
  replacing its fitting algorithm with a multistage method that combines
  local and global optimization,
\item
  supporting squares and rectangles, and
\item
  providing analytical gradients.
\end{itemize}

Eunoia serves researchers and others who need trustworthy proportional
set visualizations in R, Python, Julia, JavaScript, Rust, or C.
Developers can also embed its diagram fitter and visualizer in their own
tools, including interactive web visualizations.

\section{State of the Field}\label{state-of-the-field}

\autoref{tab:packages} compares Eunoia with representative
area-proportional diagram fitters:
venneuler~(\citeproc{ref-wilkinson2012}{Wilkinson, 2012}),
eulerAPE~(\citeproc{ref-micallef2014}{Micallef \& Rodgers, 2014}),
venn.js~(\citeproc{ref-frederickson2015}{Frederickson, 2015}),
matplotlib-venn~(\citeproc{ref-tretyakov2024}{Tretyakov, 2024}),
nVenn~(\citeproc{ref-perezsilva2018}{Pérez-Silva et al., 2018}),
Edeap~(\citeproc{ref-wybrow2021}{Wybrow et al., 2021}), and
eulerr~(\citeproc{ref-larsson2018}{Larsson \& Gustafsson, 2018}).

\begin{longtable}[]{@{}
  >{\raggedright\arraybackslash}p{(\columnwidth - 8\tabcolsep) * \real{0.2472}}
  >{\raggedright\arraybackslash}p{(\columnwidth - 8\tabcolsep) * \real{0.2472}}
  >{\raggedright\arraybackslash}p{(\columnwidth - 8\tabcolsep) * \real{0.2360}}
  >{\centering\arraybackslash}p{(\columnwidth - 8\tabcolsep) * \real{0.0562}}
  >{\raggedright\arraybackslash}p{(\columnwidth - 8\tabcolsep) * \real{0.2135}}@{}}
\caption{Related software for area-proportional Euler and Venn diagrams:
fitting algorithm, supported shapes, maximum number of sets, and
implementation language. \label{tab:packages}}\tabularnewline
\toprule\noalign{}
\begin{minipage}[b]{\linewidth}\raggedright
Package
\end{minipage} & \begin{minipage}[b]{\linewidth}\raggedright
Algorithm
\end{minipage} & \begin{minipage}[b]{\linewidth}\raggedright
Shapes
\end{minipage} & \begin{minipage}[b]{\linewidth}\centering
Sets
\end{minipage} & \begin{minipage}[b]{\linewidth}\raggedright
Language
\end{minipage} \\
\midrule\noalign{}
\endfirsthead
\toprule\noalign{}
\begin{minipage}[b]{\linewidth}\raggedright
Package
\end{minipage} & \begin{minipage}[b]{\linewidth}\raggedright
Algorithm
\end{minipage} & \begin{minipage}[b]{\linewidth}\raggedright
Shapes
\end{minipage} & \begin{minipage}[b]{\linewidth}\centering
Sets
\end{minipage} & \begin{minipage}[b]{\linewidth}\raggedright
Language
\end{minipage} \\
\midrule\noalign{}
\endhead
\bottomrule\noalign{}
\endlastfoot
\texttt{venneuler} & Stress minimization & Circles & Any &
\texttt{Java}, \texttt{R} \\
\texttt{eulerAPE} & Hill climbing & Ellipses & 3 & \texttt{Java} \\
\texttt{venn.js} & MDS + final layout & Circles & Any &
\texttt{JavaScript} \\
\texttt{matplotlib-venn} & Analytic placement & Circles & 3 &
\texttt{Python} \\
\texttt{nVenn} & Physics-based simulation & Polygons & Any &
\texttt{C++}, \texttt{R} \\
\texttt{Edeap} & Hill climbing & Ellipses & Any & \texttt{JavaScript} \\
\textbf{\texttt{Eunoia}} & MDS + final layout & Circles, ellipses,
squares, rectangles & Any & \texttt{Rust}, \texttt{R}, \texttt{Julia},
\texttt{Python}, \texttt{C}, \texttt{JS} \\
\end{longtable}

Eunoia is a separate library because eulerr's C++ implementation was
coupled to R and could not directly support WebAssembly and the other
bindings. Centralizing fitting, diagnostics, and geometry in an
independent Rust crate keeps the interfaces consistent and avoids
duplicating this work across bindings.

\section{Example}\label{example}

A diagram specification names the sets and their intersections and gives
the size of each region. For example, in Rust:

\begin{Shaded}
\begin{Highlighting}[]
\KeywordTok{use} \PreprocessorTok{eunoia::geometry::shapes::}\NormalTok{Ellipse}\OperatorTok{;}
\KeywordTok{use} \PreprocessorTok{eunoia::}\OperatorTok{\{}\NormalTok{DiagramSpecBuilder}\OperatorTok{,}\NormalTok{ Fitter}\OperatorTok{,}\NormalTok{ InputType}\OperatorTok{\};}

\KeywordTok{let}\NormalTok{ spec }\OperatorTok{=} \PreprocessorTok{DiagramSpecBuilder::}\NormalTok{new()}
    \OperatorTok{.}\NormalTok{set(}\StringTok{"SE"}\OperatorTok{,} \DecValTok{13.0}\NormalTok{)}
    \OperatorTok{.}\NormalTok{set(}\StringTok{"Treat"}\OperatorTok{,} \DecValTok{28.0}\NormalTok{)}
    \OperatorTok{.}\NormalTok{set(}\StringTok{"Anti{-}CCP"}\OperatorTok{,} \DecValTok{101.0}\NormalTok{)}
    \OperatorTok{.}\NormalTok{set(}\StringTok{"DAS28"}\OperatorTok{,} \DecValTok{91.0}\NormalTok{)}
    \OperatorTok{.}\NormalTok{intersection(}\OperatorTok{\&}\NormalTok{[}\StringTok{"SE"}\OperatorTok{,} \StringTok{"Treat"}\NormalTok{]}\OperatorTok{,} \DecValTok{1.0}\NormalTok{)}
    \OperatorTok{.}\NormalTok{intersection(}\OperatorTok{\&}\NormalTok{[}\StringTok{"SE"}\OperatorTok{,} \StringTok{"DAS28"}\NormalTok{]}\OperatorTok{,} \DecValTok{14.0}\NormalTok{)}
    \OperatorTok{.}\NormalTok{intersection(}\OperatorTok{\&}\NormalTok{[}\StringTok{"Treat"}\OperatorTok{,} \StringTok{"Anti{-}CCP"}\NormalTok{]}\OperatorTok{,} \DecValTok{6.0}\NormalTok{)}
    \OperatorTok{.}\NormalTok{intersection(}\OperatorTok{\&}\NormalTok{[}\StringTok{"SE"}\OperatorTok{,} \StringTok{"Anti{-}CCP"}\OperatorTok{,} \StringTok{"DAS28"}\NormalTok{]}\OperatorTok{,} \DecValTok{1.0}\NormalTok{)}
    \OperatorTok{.}\NormalTok{input\_type(}\PreprocessorTok{InputType::}\NormalTok{Exclusive)}
    \OperatorTok{.}\NormalTok{build()}
    \OperatorTok{.}\NormalTok{unwrap()}\OperatorTok{;}

\KeywordTok{let}\NormalTok{ layout }\OperatorTok{=} \PreprocessorTok{Fitter::}\OperatorTok{\textless{}}\NormalTok{Ellipse}\OperatorTok{\textgreater{}}\PreprocessorTok{::}\NormalTok{new(}\OperatorTok{\&}\NormalTok{spec)}\OperatorTok{.}\NormalTok{seed(}\DecValTok{1}\NormalTok{)}\OperatorTok{.}\NormalTok{fit()}\OperatorTok{.}\NormalTok{unwrap()}\OperatorTok{;}
\end{Highlighting}
\end{Shaded}

\begin{figure}
\centering
\includegraphics{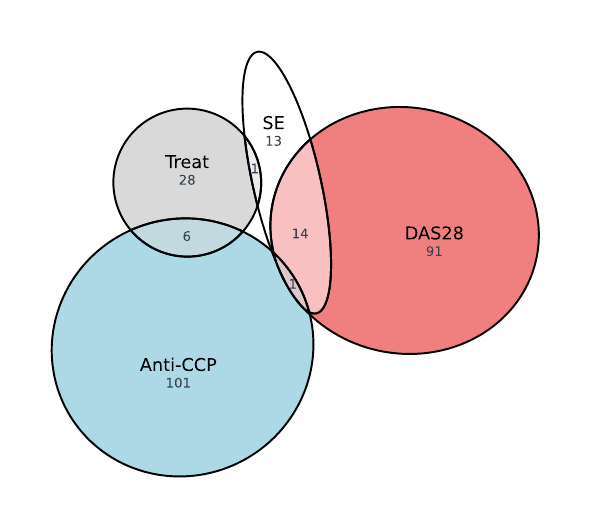}
\caption{A four-set Euler diagram fitted with ellipses, with region
quantities shown. Eunoia fits it to a diagError of \(10^{-4}\), an
essentially exact diagram.\label{fig:ellipse}}
\end{figure}

The four sets---SE, Treat, Anti-CCP, and DAS28---correspond to clinical
variables in the rheumatoid arthritis data from Junta et al.
(\citeproc{ref-junta2009}{2009}). The fitted layout is shown in
\autoref{fig:ellipse}. Its diagError of \(10^{-4}\) indicates that the
diagram reproduces the data essentially exactly. The figure also shows
Eunoia's automatic label placement: set labels and region quantities are
positioned at poles of inaccessibility of the region polygons, with
labels placed outside the diagram when they do not fit inside their
regions.

The bindings mirror this interface; the equivalent call in Python is:

\begin{Shaded}
\begin{Highlighting}[]
\ImportTok{import}\NormalTok{ eunoia }\ImportTok{as}\NormalTok{ eu}

\NormalTok{diagram }\OperatorTok{=}\NormalTok{ eu.euler(}
\NormalTok{    \{}
        \StringTok{"SE"}\NormalTok{: }\DecValTok{13}\NormalTok{,}
        \StringTok{"Treat"}\NormalTok{: }\DecValTok{28}\NormalTok{,}
        \StringTok{"Anti{-}CCP"}\NormalTok{: }\DecValTok{101}\NormalTok{,}
        \StringTok{"DAS28"}\NormalTok{: }\DecValTok{91}\NormalTok{,}
        \StringTok{"SE\&Treat"}\NormalTok{: }\DecValTok{1}\NormalTok{,}
        \StringTok{"SE\&DAS28"}\NormalTok{: }\DecValTok{14}\NormalTok{,}
        \StringTok{"Treat\&Anti{-}CCP"}\NormalTok{: }\DecValTok{6}\NormalTok{,}
        \StringTok{"SE\&Anti{-}CCP\&DAS28"}\NormalTok{: }\DecValTok{1}\NormalTok{,}
\NormalTok{    \},}
\NormalTok{    shape}\OperatorTok{=}\StringTok{"ellipse"}\NormalTok{,}
\NormalTok{)}
\end{Highlighting}
\end{Shaded}

The JavaScript, Julia, and R packages follow the same pattern.

\begin{figure}
\centering
\includegraphics{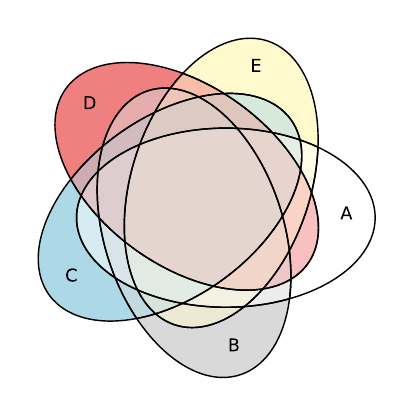}
\caption{A canonical five-set Venn diagram drawn with
ellipses.\label{fig:venn}}
\end{figure}

Eunoia also constructs Venn diagrams, in which every intersection is
drawn regardless of whether it is empty~(\autoref{fig:venn}).

\section{Software Design}\label{software-design}

\begin{figure}
\centering
\includegraphics{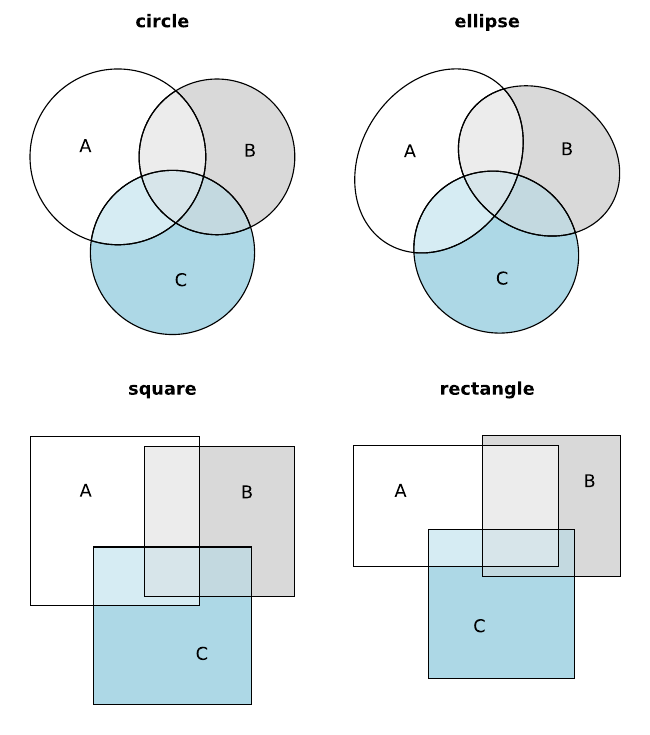}
\caption{A three-set specification fitted with circles, ellipses,
squares, and rectangles. Not every shape family can represent the input
exactly: the diagError is \(1.2 \times 10^{-2}\) for circles,
\(1.7 \times 10^{-2}\) for squares, and \(4.5 \times 10^{-2}\) for
rectangles, compared to \(2.3 \times 10^{-14}\) for
ellipses.\label{fig:shapes}}
\end{figure}

The library uses a trait-based design, with shapes selected at compile
time. A given specification can be fitted with circles, ellipses,
squares, or rectangles, as illustrated in \autoref{fig:shapes}.

Fitting proceeds in two phases, roughly following the approach of
venneuler~(\citeproc{ref-wilkinson2012}{Wilkinson, 2012}),
venn.js~(\citeproc{ref-frederickson2015}{Frederickson, 2015}), and the
original eulerr package~(\citeproc{ref-larsson2018}{Larsson \&
Gustafsson, 2018}). The first phase uses multidimensional scaling to
place shapes of fixed size so that their pairwise distances approximate
those required by the pairwise intersections. For disjoint and contained
set pairs, it uses the relaxed loss suggested by Frederickson
(\citeproc{ref-frederickson2015}{2015}). The second phase refines all
shape parameters to minimize a loss computed from the differences
between the fitted and requested region areas\footnote{The default loss
  is a normalized sum of squared errors.}. Eunoia computes region areas
analytically for every shape. It finds ellipse intersections through a
projective-conic
construction~(\citeproc{ref-richtergebert2011}{Richter-Gebert, 2011})
and computes the resulting overlaps from circular and elliptical
segments in closed form~(\citeproc{ref-eberly2016}{Eberly, 2016}). For
several shape families, the smooth losses have exact analytical
gradients.

For shapes with analytical gradients, refinement begins with
Levenberg-Marquardt least squares, with a bounded Levenberg-Marquardt
retry if the loss remains above a threshold. If both local fits remain
above the threshold, a bounded variant of
CMA-ES~(\citeproc{ref-hansen1996}{Hansen \& Ostermeier, 1996}) searches
for another basin, which bounded Levenberg-Marquardt then polishes.
Eunoia keeps the solution with the lowest loss. The pipeline uses
several seeded restarts, which can run in parallel when the optional
\texttt{parallel} feature is enabled (but remain serial in WebAssembly).
For small set counts, one restart begins from a canonical Venn layout.
Non-smooth losses---such as the maximum absolute region error---use
derivative-free methods (Nelder--Mead or mesh-adaptive direct search) or
can be replaced by a smooth surrogate.

Eunoia returns the fitted layout with its residuals, loss, and two
goodness-of-fit statistics: stress, a venneuler-style normalized
least-squares measure, and diagError, the maximum absolute difference
between requested and fitted region shares. These diagnostics allow
users to assess the diagram's quality numerically. A plotting module
extracts the region polygons through polygon clipping and computes label
positions. The R, Python, Julia, and JavaScript packages use this output
for rendering. The core has no dependency on any host language's runtime
and compiles to WebAssembly. The \href{https://eunoia.bz/app}{web app}
therefore fits diagrams entirely in the browser without a server.

\section{Annotation and
Interactivity}\label{annotation-and-interactivity}

\begin{figure}
\centering
\includegraphics{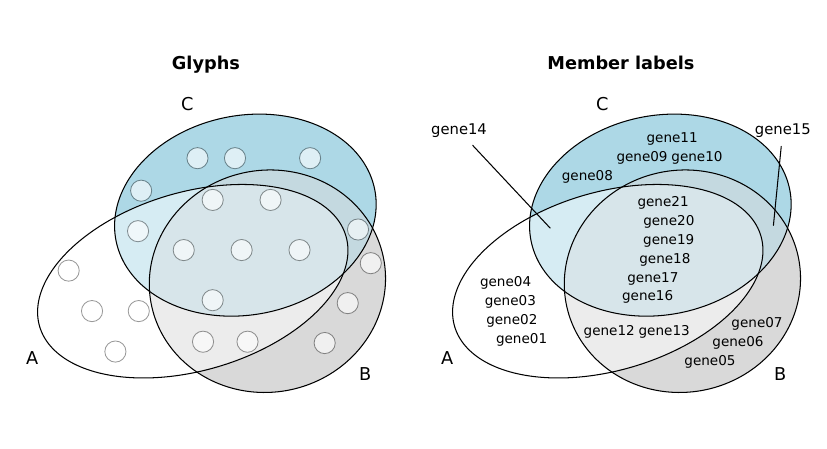}
\caption{A three-set Euler diagram fitted with ellipses, shown with
glyphs (left) and member labels (right). Callouts identify members in
regions too narrow for their names.\label{fig:features}}
\end{figure}

Area-proportional Euler diagrams encode the size of each set
intersection as the area of a region. Eunoia adds three complementary
features: unit glyphs show how many elements a region contains, member
labels identify those elements, and interactive output reveals
information about regions too small to annotate
(\autoref{fig:features}).

\subsection{Glyphs}\label{glyphs}

Glyphs are circular marks, one per data unit, packed inside their
regions following the eulerGlyphs approach of Micallef et al.
(\citeproc{ref-micallef2012}{2012}). They can also be added to Venn
diagrams. All glyphs in a diagram are the same size, so counts can be
compared across regions. By default, Eunoia searches for the largest
size that fits all the glyphs in every region. The placement routines
also accept a fixed size and report any glyphs that do not fit. Glyphs
can be kept clear of the diagram's labels where space allows.

The shared placement routines offer four arrangements: a hexagonal grid
(the default), a rectangular grid, an even, deterministic scatter, and a
random scatter.

\subsection{Member Labels}\label{member-labels}

Member labels show the names of the elements in each region. The shared
packing routine places each name individually inside its region and
reduces all font sizes by the same factor when space is limited, down to
a minimum size. Names that still do not fit are reported as unplaced.

\subsection{Interactivity}\label{interactivity}

The Python package provides an interactive Plotly backend through
\texttt{fit.plot\_plotly()}. Each region has a hover tooltip showing its
intersection label, sample size, and members. These tooltips are
primarily useful for dense multi-set diagrams, in which small regions
cannot hold any text. The JavaScript SVG serializer similarly supports
native tooltips and data attributes on each region, so host pages can
add their own interaction. The web app provides controls for glyph mode,
spacing, and styling.

\section{Research Impact Statement}\label{research-impact-statement}

Eunoia descends from eulerr, which has been distributed on CRAN since
2016 and has been cited in over 700 academic publications\footnote{Google
  Scholar, accessed Aug 11, 2026}, predominantly in the life sciences.
Since version~8.0, eulerr has used Eunoia in place of its C++ backend,
so all current eulerr users also use Eunoia. Other packages depend on
eulerr, and therefore on Eunoia, to draw area-proportional diagrams in
their analyses. These include three Bioconductor genomics
packages---cola, hicVennDiagram, and seqsetvis---and the CRAN package
RulesTools. Six more packages---DOTSeq, highdir,
IlluminaHumanMethylationEPICv2manifest, ISAnalytics, overviewR, and
pcutils---use eulerr optionally and can therefore also invoke Eunoia.

The Python, Julia, and JavaScript packages and the web app extend the
same underlying library to communities that previously had no access to
ellipse-based area-proportional diagrams.

\section{AI Usage Disclosure}\label{ai-usage-disclosure}

JL wrote Eunoia's predecessor, eulerr, entirely without AI. For Eunoia,
JL used generative AI tools including Claude Code Opus (versions 4.5,
4.8, and 5.0), Claude Code Fable 5, GitHub Copilot GPT 5.1 and 5.4, and
Codex GPT 5.6-sol and GPT 6.1-sol. NML used Claude Code Opus 5.5. The
authors used these tools to help write code, parts of the documentation,
and unit tests and to review and edit the manuscript. NML performed user
testing and detected bugs independently of AI. JL and NML reviewed,
modified, and validated all AI-assisted content and made the final
design and implementation decisions.

\section{Acknowledgements}\label{acknowledgements}

Eunoia builds on a decade of feedback from users of eulerr, whose bug
reports and feature requests have shaped the fitting engine, and on
Peter Gustafsson's contributions to the original eulerr package.

\section*{References}\label{references}
\addcontentsline{toc}{section}{References}

\phantomsection\label{refs}

\end{document}